\documentclass[%
 reprint,
 amsmath,amssymb,
 aps,prl,
]{revtex4-2}

\usepackage{graphicx}
\usepackage{dcolumn}
\usepackage{bm}

\begin{document}

\preprint{APS/123-QED}

\title{Dissipative coupling induced UWB magnonic frequency combs generation}

\author{Zeng-Xing Liu}
 \email{zengxingliu@hust.edu.cn}
\affiliation{School of Electronic Engineering $\&$ Intelligentization, Dongguan University of Technology, Dongguan, Guangdong 523808, China
}%


\date{\today}

\begin{abstract}
Magnonic frequency combs have recently attracted particular attention due to their potential impact on spin-wave science.
Here, we demonstrate theoretically the generation of ultra-wideband (UWB) magnonic frequency combs induced by dissipative coupling in an open cavity magnomechanical system.
A broadband comb with gigahertz repetition rates is obtained in the magnonic spectrum and a typical non-perturbation frequency-comb structure is also observed.
The total width of the magnonic comb in the robust plateau region can be up to $\sim400$ comb lines, which is much broader and flatter than the reported in the previous works.
Furthermore, when the dissipative coupling strength is further increased, the chaotic motion is predicted in the magnonic spectrum.
Our results provide an in-depth understanding of nonlinear magnomechanic dynamics in open quantum systems and fundamentally broadens the research range of magnon in wider spectral regimes.
\end{abstract}

\maketitle


Magnons, the collective excitations of spin waves in magnetic materials, have attracted attention due to their excellent physical properties and potential applications in information processing \cite{YIG,spintronics,Current,Current1,Insulators2}.
Magnons and their related quantum systems are rich in nonlinearity, and many interesting magnonic phenomena have been observed both in classical and quantum regimes \cite{Quantum,Nonlinear,Hybrid,Ultra,Hybrid1,arXiv}.
Recently, the concept of magnonic frequency comb has been proposed \cite{comb1,comb2,comb3,comb4,comb5,comb6,comb7,comb8}, which can be regarded as a magnonic counterpart to the optical frequency comb.
The extension of frequency comb techniques to the field of spin waves is likely to open up attractive prospects for applications, especially in spectroscopy and magnon-based precision metrology \cite{Insulators0,Insulators,Insulators1}.
However, due to the intrinsically weak magnonic nonlinearity, it is difficult to generate UWB magnonic frequency combs with a robust plateau region.
To this aim, the generation of magnonic frequency comb based on resonance enhanced magnetostrictive interaction has been proposed theoretically \cite{comb3} and quickly verified experimentally \cite{comb6}.
Furthermore, the efficiency of magnonic frequency comb generation can also be enhanced via a two-tone microwave drive \cite{comb8}, and the use of exceptional points in a coupled magnonic system has also been shown to generate magnonic frequency combs efficiently \cite{comb7}.
Nevertheless, the weak tunability and poor bandwidth (just a few pronounced comb lines) are prominent shortcomings that need to be solved urgently.


In the present work, we propose an efficient mechanism to induce UWB magnonic frequency combs generation by introducing a dissipative coupling in an open cavity magnomechanical system.
The system consists of three different modes, namely the magnon mode of the Yttrium Iron Garnet (YIG) sphere, the microwave cavity mode and the mechanical mode, in which the magnon and microwave cavity mode interact with each other through dissipative coupling, and the magnon and mechanical modes interact via magnetostriction.
Our results show that when the system is dominated by dissipative coupling, a distinct plateau region and a cutoff region appear in the magnonic spectrum.
Advantageously, the total width of the magnonic frequency comb in the plateau region can be up to $\sim400$ comb lines with a span of about 28.7 GHz, which is wider and flatter than what has been reported in the previous literatures \cite{comb1,comb2,comb3,comb4,comb5,comb6,comb7,comb8}.
Furthermore, abundant non-perturbation signals are observed in the magnonic spectrum, which is similar to the high-order harmonic generation in atom-molecular systems \cite{harmonic}.
In addition, the generation of magnonic frequency combs can be controlled by precisely moving the YIG sphere.
Our results suggest the open cavity magnomechanical system mediated by dissipative coupling can be a promising platform for investigating UWB magnonic frequency combs, and brings potential application for magnonic frequency conversion and precision calibration \cite{Insulators0,Insulators,Insulators1}.

\begin{figure}[htb]
\centering\includegraphics [width=1\linewidth] {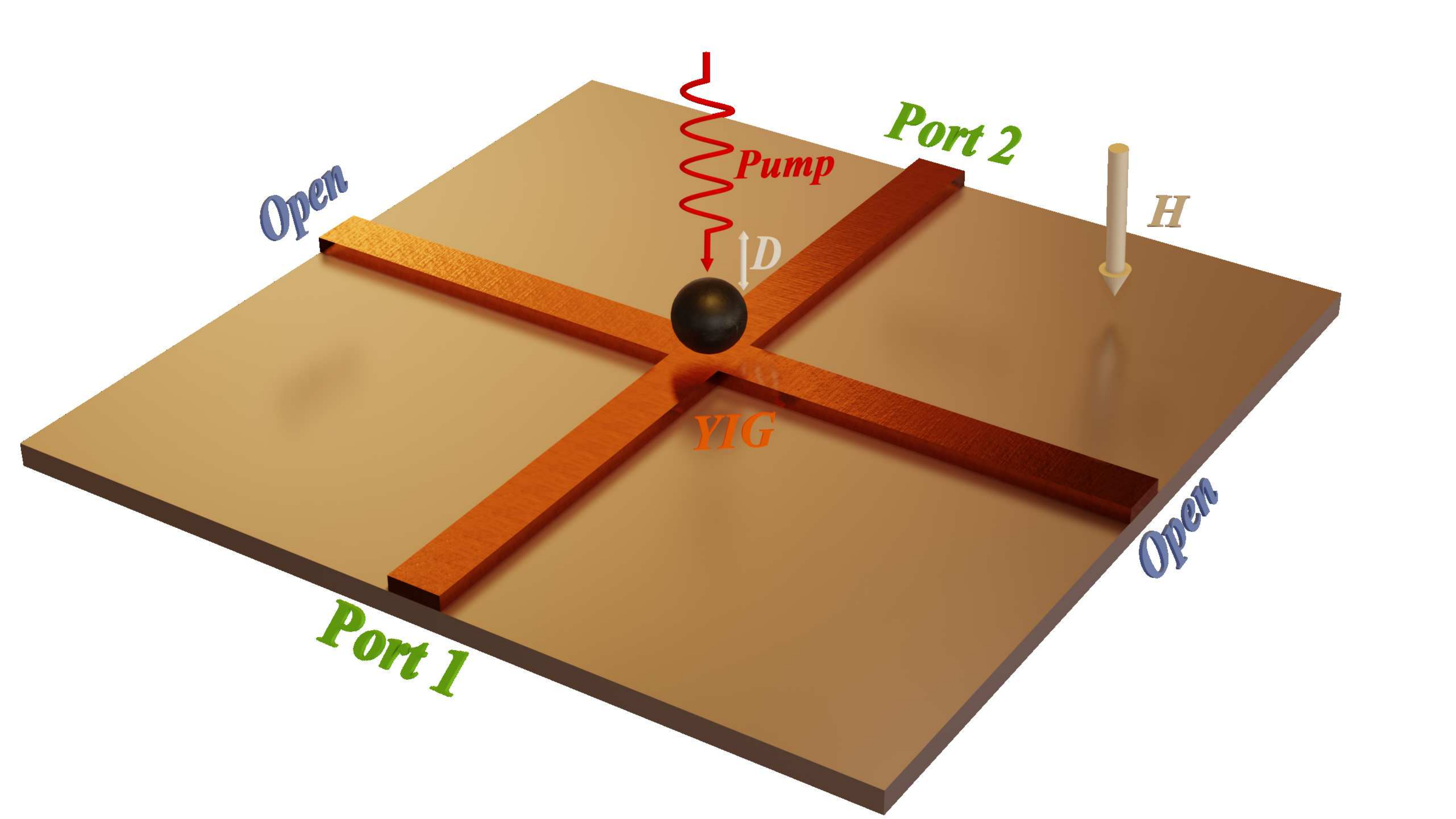}
\caption{(a) Schematic illustration of an open cavity magnomechanical system, in which a YIG sphere is placed above the center of the crossline-microwave circuit that supports a traveling wave inducing the dissipative coupling between the cavity and magnon modes.
The YIG sphere is driven by a microwave field, and the probe of the system can be achieved through ports 1 and 2.
A uniform magnetic field (H) is applied along the $z$ direction to saturate the magnetization.}
\label{fig:0}
\end{figure}

We consider an open cavity magnomechanical system, as schematically shown in Fig. \ref{fig:0}(a), in which a highly polished single-crystal YIG sphere is placed above the center of the crossline-microwave circuit.
The crossline-microwave circuit supports both the standing wave that forms the cavity mode and the traveling wave that induces the dissipative coupling between the cavity mode and the Kittel mode by radiating the energy to open environment, which has been proposed theoretically and demonstrated experimentally \cite{dissipative3,mechanism,dissipative4,dissipative1,dissipative2}.
A uniform magnetic field (with the strength H) is applied along the $z$ direction to saturate the magnetization, and the frequency of the Kittel mode is controlled by the external bias magnetic field, i.e., $\omega_m = \varrho \rm{H}$, where $\varrho/2\pi = 28$ $\rm{GHz/T}$ is the gyromagnetic ratio \cite{YIG}.
The Hamiltonian including the microwave cavity mode, the Kittel mode, and their dissipative interactions can be described in the form ($\hbar\equiv1$) $H_{0} = (\omega_{a}-i\kappa_{a})\hat{a}^{\dagger}\hat{a}+(\omega_{m}-i\kappa_{m})\hat{m}^{\dagger}\hat{m}
-iJ_{am}(\hat{a}^{\dagger}\hat{m}+\hat{a}\hat{m}^{\dagger})$ \cite{dissipative1}, where $\hat{a} (\hat{a}^{\dagger})$ is the boson annihilation (creation) operator of the cavity mode (with the intrinsic frequency $\omega_{a}$ and the damping rate $\kappa_{a}$).
$\hat{m} (\hat{m}^{\dagger})$ is the boson annihilation (creation) operator of the Kittel mode (with the intrinsic frequency $\omega_{m}$ and the damping rate $\kappa_{m}$).
$J_{am}$ is the dissipative coupling rate between the cavity and the Kittel modes, which can be changed when moving the YIG sphere \cite{dissipative1,dissipative2}.
It should be noted that coherent magnon-photon coupling and dissipative magnon-photon coupling exist simultaneously and compete with each other. Here, we discussed the case that when the dissipative coupling is dominant, because the change of coherent coupling strength has little effect on the magnonic frequency combs generation.
The YIG sphere is directly driven by a microwave drive field with the driving frequency $\omega_{l}$, and the Hamiltonian can be written as ${H_{d}} =i\Omega_{0}(\hat{m}^{\dagger}e^{-i\omega_{l}t}-\hat{m}e^{i\omega_{l}t})$ under the low-lying excitation approximation, i.e., the excited magnon number is much smaller than the total spin number $\langle m^{\dagger}m\rangle/2N_{\rm{spin}}\ll 1$.
Here, $\Omega_{0}=\sqrt{5N_{\rm{spin}}}\varrho B_{0}/4$ is the Rabi frequency of the microwave drive field with $B_{0}$ the driving amplitude, $N_{\rm{spin}}$ the total number of the spins in the YIG sphere.
According to the magnetostrictive effect, varying magnetization induced by the excitation of magnons inside the YIG sphere leads to the deformation of its spherical geometry, and vice versa, resulting in coherent coupling between the Kittel mode and the vibrational mode \cite{phonon1}.
In this case, the magnetostrictive interaction between the Kittel and the vibrational modes can be depicted by a radiation pressure-like Hamiltonian, i.e., ${H_{int}} =\hbar g_{mb}\hat{m}^{\dagger}\hat{m}(\hat{b}^{\dagger}+\hat{b})$, where $\hat{b} (\hat{b}^{\dagger})$ is the boson annihilation (creation) operator of the vibrational mode (with the intrinsic frequency $\omega_{b}$ and the damping rate $\kappa_{b}$), and $g_{mb}$ is the single magnon-phonon coupling strength \cite{phonon1,phonon2,Magnetostriction2,Magnetostriction1}.

In order to make the Hamiltonian time independent, a unitary transformation $\mathbf{U}(t) = {\rm{\exp}}(-i\omega_{l}\hat{m}^{\dagger}\hat{m}t)$ is applied. Thus, we have
\begin{eqnarray}\label{equ:8}
  \rm{H} &=& \mathbf{U}(t)H\mathbf{U}^{\dagger}(t)-
  i\hbar\mathbf{U}(t)\frac{\partial\mathbf{U}^{\dagger}(t)}{\partial t}\nonumber \\
  &=&\hbar\Delta_{a}\hat{a}^{\dagger}\hat{a}+\hbar\Delta_{m}\hat{m}^{\dagger}\hat{m}
  -iJ_{am}(\hat{a}^{\dagger}\hat{m}+\hat{a}\hat{m}^{\dagger})\nonumber \\
  &&+\hbar\omega_{b}\hat{b}^{\dagger}\hat{b}+
  \hbar g_{mb}\hat{m}^{\dagger}\hat{m}(\hat{b}^{\dagger}+\hat{b})
  +i\hbar\Omega_{0}(\hat{m}+\hat{m}^{\dagger}),
\end{eqnarray}
where $\Delta_{a(m)}=\omega_{l}-\omega_{a(m)}$ is the detuning from the microwave pumping field and the cavity (magnon) mode in a frame rotating at $\omega_{l}$.
Considering the system dissipation with the Markov approximation \cite{Noise}, the evolution of each mode of the system can be described by the Heisenberg-Langevin equation, i.e., $\dot{\hat{o}} = i/\hbar[{{H}},\hat{o}] (\hat{o}=\hat{b}, \hat{a}, \hat{m}$), read as
\begin{eqnarray}\label{equ:1}
\frac{d}{dt}b&=& (-i\omega_{b}-\kappa_{b})b-ig_{mb}m^{\ast}m, \nonumber \\
\frac{d}{dt}a&=& (i\Delta_{a}-\kappa_{a})a-\Gamma m,  \\
\frac{d}{dt}m&=& (i\Delta_{m}-\kappa_{m})m-\Gamma a- ig_{mb}(b+b^{\ast})m+\Omega.\nonumber
\end{eqnarray}
Under the semiclassical approximation, the operators of each mode are reduced to their average values, viz. $a(t)\equiv \hat{a}(t)$, $m(t)\equiv \hat{m}(t)$, and $b(t)\equiv \hat{b}(t)$ \cite{Noise}.
The thermal noise terms are neglected safely due to the strong robustness of the magnonic frequency comb generation to thermal noise, which is demonstrated to be valid in Ref \cite{comb3,comb8}.
Equation (\ref{equ:1}) is a set of coupled mode equations whose self-consistent analytic solution can be mathematically obtained by using the ansatz $b=\sum_{\ell=-\infty}^{\infty}b_{\ell}e^{-i\ell\omega_{b}t}$, $a=\sum_{\jmath=-\infty}^{\infty}a_{\jmath}e^{-i(\omega_{l}+\jmath\omega_{b}t)}$, and $m=\sum_{n=-\infty}^{\infty}m_{n}e^{-i(\omega_{l}+n\omega_{b}t)}$, where $b_{\ell}$, $a_{\jmath}$, and $m_{n}$ are complex amplitudes of the mechanical, microwave cavity, and magnon harmonics with $\ell$, $\jmath$, and $n$ being integers representing the harmonic indexes of the corresponding modes, respectively.
Physically, the YIG sphere is directly driven via a local microwave antenna to excite the magnon mode, and on the other hand, the change of magnetization cause deformation of the YIG sphere, such that the vibrational mode can be excited.
Then the vibrational mode can couple with the driving mode and excite the sum-frequency ($\omega_{l}+\omega_{b}$) and difference-frequency ($\omega_{l}-\omega_{b}$) modes, while these excitation modes can further couple with the vibrational mode to generate higher-order frequency modes \cite{Conversion2}. 
Thus, spectral components with a series of equally spaced teeth around the pumped mode $\omega_{l}$, i.e., $\omega = \omega_{l} \pm n\omega_{f}$ with $\omega_{f} = \omega_{b}$ the repetition frequency, will appear in the magnonic spectrum, which is the main characteristic of the frequency combs \cite{Insulators0}.


\begin{figure}[htb]
\centering
\includegraphics [width=1\linewidth] {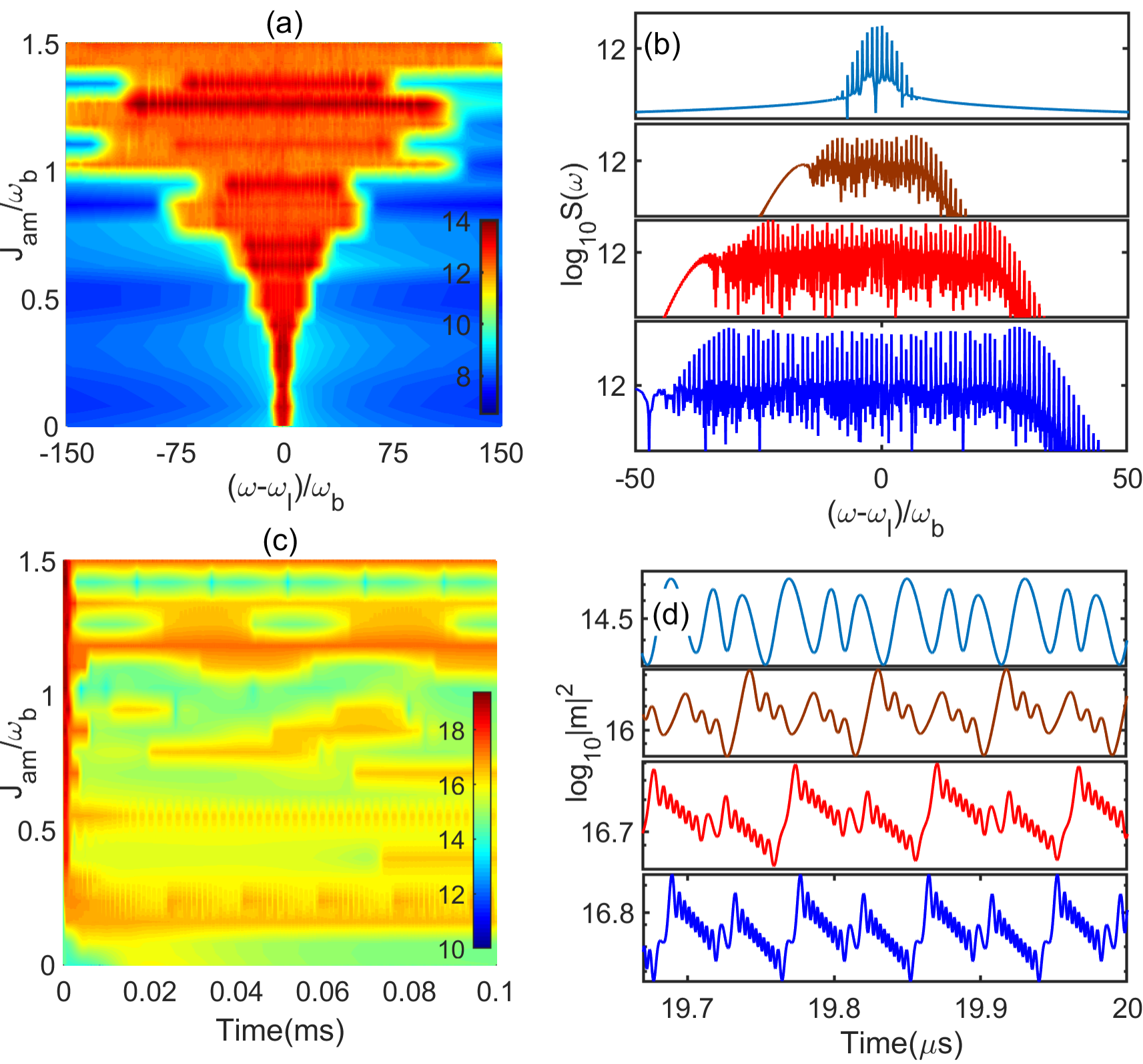}
\caption{(a) The contour plot of the magnonic frequency comb spectrum with different dissipative coupling strengths $J_{am}$.
The color bar represents the amplitude of the frequency comb in logarithmic scale $\log_{10}{\rm{S}}(\omega)$.
(b) The magnonic frequency comb spectra under different dissipative coupling strengths $J_{am}/\omega_{b}=0, 0.5, 0.8, 1$, respectively.
(c) Time evolution of the magnon number $|m|^2$ varies with dissipative coupling strengths.
The color bar represents the magnon number in logarithmic scale $\log_{10}|m|^2$.
(d) The magnon number under different dissipative coupling strengths $J_{am}/\omega_{b}=0, 0.5, 0.8, 1$ in stable states, respectively.
The parameters of the magnomechanical system are chosen from the recent experiment \cite{phonon1,dissipative1} $\omega_{b}/2\pi=11.42$ $\rm{MHz}$, $\kappa_{m}/2\pi=0.56$ $\rm{MHz}$, $\kappa_{b}/2\pi=150$ $\rm{Hz}$, $g_{mb}/2\pi=9.88$ $\rm{mHz}$, $\Delta_{a} = \Delta_{m}=\omega_{b}$, and the microwave drive field amplitude $B_{0}=3.5\times10^{-5}$ T.}
\label{fig:1}
\end{figure}

In the present work, Eq.\ref{equ:1} is solved numerically by using the Runge-Kutta method.
The number of steady-state modes is much greater than the number of modes at the initial moment, so for convenience, the initial conditions are chosen to $b|_{t=0}=0$, $a|_{t=0}=0$, and $m|_{t=0}=0$.
Further, the magnonic spectrum can be obtained by doing the fast Fourier transform of the time series \cite{FFT}.
Figure \ref{fig:1}(a) shows the magnonic spectrum varies with the dissipative coupling strength between the cavity and the Kittel modes.
The result shows that when the dissipative coupling strength is relatively weak, the generation efficiency of the magnonic frequency comb undergoes a slow enhancement process, and with the increasing of the dissipative coupling strength, the generation of the magnonic frequency comb is dramatically enhanced from about $J_{am}/\omega_{b}\approx0.5$.
More specifically, in the absence of dissipative coupling, i.e., $J_{am}/\omega_{b}=0$, as shown by the sky-blue line in Fig. \ref{fig:1}(b), a few comb teeth can be observed in the magnonic spectrum due to the resonantly enhanced magnetostrictive effect, which has been discussed in the latest Ref.\cite{comb3}.
Nevertheless, magnonic frequency combs generated based on this mechanism can hardly obtain an ultra-wideband spectrum, and it is difficult to observe typical nonperturbative spectral structure, such as plateau and cutoff.
To solve this bottleneck, we propose an another mechanism to induce UWB magnonic frequency combs generation by introducing a dissipative coupling in an open cavity magnomechanical system.
Advantageously, when the dissipative coupling strength is taken to be $J_{am}/\omega_{b}=0.5, 0.8, 1$, we observe a high dependence of the magnonic spectrum on the dissipative coupling strength with the same experimental parameters, and the results are shown in the brown, red and blue lines in Fig. \ref{fig:1}(b).
Compared to previous schemes \cite{comb1,comb2,comb3,comb4,comb5,comb6,comb7,comb8}, the dissipative-induced magnonic frequency comb spectra have the characteristic structure of ultra-wideband, robust plateau and cutoff, which is quite similar to high order harmonic generation in atom-molecular systems \cite{harmonic}.
Furthermore, the time evolution of the magnon number on the dissipative coupling strength is shown in Fig. \ref{fig:1}(c).
Under the action of dissipative coupling, the magnonic dynamics undergoes a process where the magnon number grows rapidly over time, and finally tends to be stable.
The steady-state magnon number is a strong evidence reflecting the nonlinear strength of the system. Obviously, the dissipative coupling between the cavity and magnon modes can greatly enhance the steady-state magnon number [increase by about two orders of magnitude, as shown in Fig. \ref{fig:1}(d)], which also indicates that the magnonic nonlinearity induced by magnetostrictive interaction is enhanced, and the wave mixing process is strengthened.
Thus, a robust magnonic frequency comb is generated.

\begin{figure}[htbp]
\centering
\includegraphics [width=1\linewidth] {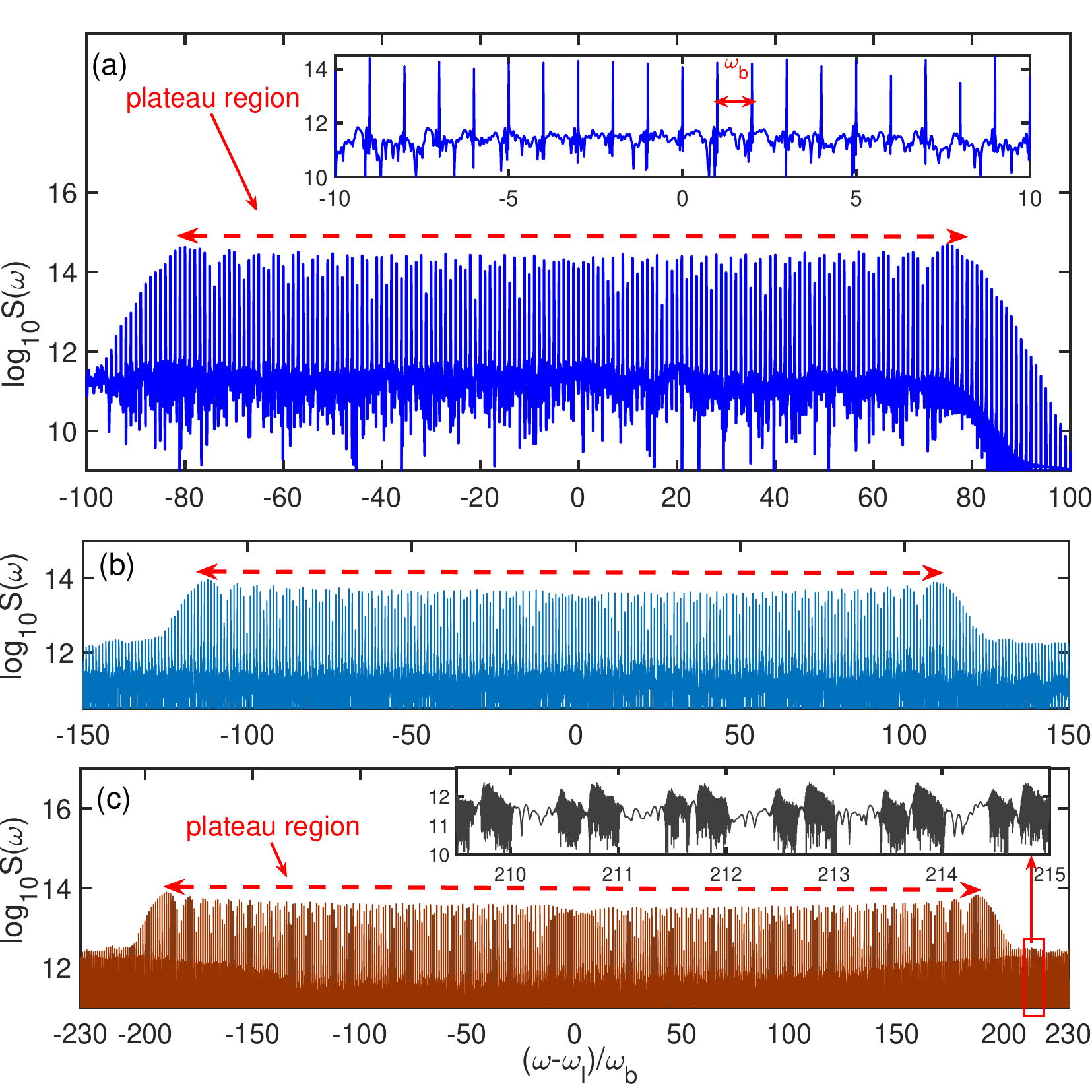}
\caption{The magnonic frequency comb spectrum under the dissipative coupling strength (a) $J_{am}/\omega_{b}=1.3$, (b) $J_{am}/\omega_{b}=2$, and (c) $J_{am}/\omega_{b}=3$.
The other parameters are the same as those in Fig.\ref{fig:1}.}
\label{fig:6}
\end{figure}

In what follows, to better understand the magnonic frequency combs, we analyze in detail the structural characteristics of magnonic spectrum under the condition of dissipative coupling strength $J_{am}/\omega_{b}=1.3$, and the result is shown in Fig. \ref{fig:6}(a).
Interestingly, a flat magnonic frequency comb with about 160 comb lines is obtained in the magnonic spectrum.
Observingly, the magnonic spectrum presents a typical frequency-comb structure [as shown in the illustration in Fig. \ref{fig:6}(a)], where each magnonic comb provides extremely narrow linewidth and excellent frequency accuracy (with fixed comb spacing).
Furthermore, a robust plateau region appears in the magnonic spectrum where all the magnonic combs have almost the same intensity [as indicated by the horizontal red dotted line in Fig. \ref{fig:6}(a)], and finally ends up with a sharp cutoff around the 80th-order tooth.
Moreover, abundant non-perturbation signals can be observed in the plateau region, for example, the amplitude of the 9th-order comb tooth is significantly greater than the 8th-order comb tooth.
To get a more robust UWB magnonic frequency comb, the dissipative coupling strengths are further increased to
$J_{am}/\omega_{b}=2$ and $J_{am}/\omega_{b}=3$, and the results are shown in Fig. \ref{fig:6} (b) and (c), respectively.
Note that the dissipative coupling strength can be controlled by precisely moving the YIG sphere in the crossline-microwave circuit \cite{dissipative1,dissipative2}, which increases the tunability of magnonic frequency comb generation.
As expected, we have achieved a UWB magnonic frequency comb that exhibits $\sim400$ comb lines with a span of about 28.7 GHz [Fig. \ref{fig:6} (c)], which is much broader and flatter than the reported in the previous works \cite{comb3,comb6,comb8}.
As the strength of the dissipative coupling increases, we also observe an interesting phenomenon, that is, the magnonic spectrum becomes more and more disordered until chaos occurs, and the result is shown in the illustration in Fig. \ref{fig:6} (c).
For one thing, the increase of dissipative coupling strength will continuously enhance the magnonic nonlinearity, for another, more energy from the outside will also bring more thermal noise, making the magnonic dynamic become more and more disorderly.
More fundamental physical processes require further analysis.

Note that although many methods have been proposed to generate magnonic frequency combs \cite{comb1,comb2,comb3,comb4,comb5,comb6,comb7,comb8}, however, all these schemes only obtain several order of magnonic combs, which seriously limits the practical application of frequency combs in the field of magnon.
Therefore, it is of great significance to propose a scheme to realize UWB magnonic frequency combs, both in physical sense and in practical application level.
Compared with previous reports, the dissipative coupling induced magnonic frequency combs generation proposed here possesses several merits: First, an ultra-bandwidth magnonic spectrum with more comb lines and a wider frequency range can be obtained, which can extend the magnonic frequency comb to a shorter spin-wave range.
Second, the repetition rate of the comb is strictly equal to the frequency of the mechanical mode, and thus all the frequency components of the magnonic frequency comb can be well defined as the calibration.
Once the center frequency of the magnonic spectrum is precisely measured, its resolution is equal to the line spacing of the magnonic frequency comb.
Finally, the physical model we discussed has a good experimental basis, and all parameters are within current experimental reach \cite{phonon1,dissipative1}.
Besides, the magnonic frequency combs based on cavity magnomechanical system are robust against the thermal noise at room temperature \cite{comb3,comb6,comb8}, consequently our scheme is highly achievable experimentally.

In short, dissipative coupling induced UWB magnonic frequency combs generation are discussed.
We show that the introduction of dissipative coupling can achieve global coupling between the excited magnon and vibration modes and excite a large number of modes, generating a UWB magnonic frequency comb.
A typical frequency-comb structure is observed in the magnonic spectrum, including a robust plateau region and cutoff, which strongly confirms that magnons may also exhibit phenomena that are similar to those in atomic-molecular systems.
Our proposal effectively solves the outstanding disadvantages of weak tunability and poor bandwidth of the magnonic frequency combs.
We preconceive that this scheme should also hold for other magnonic quantum systems since magnon possess excellent compatibility with other quasiparticles (for example, photons, qubits, and spin textures).

This work was supported by the National Science Foundation (NSF) of China (Grants No. 12105047), the Guangdong Basic and Applied Basic Research Foundation (Grant No. 2022A1515010446).

\end{document}